\documentclass[referee,sn-mathphys-num]{sn-jnl}

  \usepackage{lineno}

\usepackage{graphicx}%
\usepackage{multirow}%
\usepackage{amsmath,amssymb,amsfonts}%
\usepackage{amsthm}%
\usepackage{mathrsfs}%
\usepackage[title]{appendix}%
\usepackage{xcolor}%
\usepackage{textcomp}%
\usepackage{manyfoot}%
\usepackage{booktabs}%
\usepackage{algorithm}%
\usepackage{algorithmicx}%
\usepackage{algpseudocode}%
\usepackage{listings}%
\usepackage{placeins}

\begin{document}

\title[Inertialess turbulence in viscoelastic pipe flow]{Inertialess turbulence in viscoelastic pipe flow}
\author[1]{\fnm{Shoaib} \sur{Kamil}}

\author*[1]{\fnm{Björn} \sur{Hof}}\email{bhof@ist.ac.at}

\affil[1]{\orgname{Institute of Science and Technology Austria (ISTA)}, \orgaddress{\street{Am Campus 1}, \city{Klosterneuburg}, \postcode{3400}, \country{Austria}}}


\abstract{

Turbulence is synonymous with inertia dominated fluid motion, entailing high velocities and large spatial scales.  Conversely, in complex fluids, elastic material properties can promote and sustain turbulent-like motions even at arbitrarily small inertia. The common route to this purely elastic state of turbulence, however, strictly occurs only in specific flows with curved streamlines, thus excluding canonical cases such as pipe flow. Investigating dilute polymer solutions in pipe experiments across an unprecedented elasticity and viscosity range, we trace an instability, previously assumed to require finite inertia, to Reynolds numbers four orders of magnitude below the theoretically predicted minimum level. At sufficiently high polymer concentrations and large shear rates, an abrupt drop in the transition threshold confirms that the vanishing inertia limit has been reached, and consequently that the fluctuating motions observed are of purely elastic origin.
}

\maketitle

While solids and liquids constitute distinct phases of matter, many practically relevant liquids, e.g., polymer solutions, paints, blood, or saliva, are not simply viscous but possess certain levels of elasticity, a property conventionally associated with solids. 
This combination of fluid and solid-like attributes gives rise to a rich variety of phenomena that often defy common intuition based on ordinary fluids.  
A striking example is that ordered motion can break down and give way to turbulence even if inertia, the common cause of turbulence, is virtually absent. Accordingly, the Reynolds number, the ratio of inertial to viscous forces, can be orders of magnitude below the levels that are usually required to sustain turbulence, yet in these more complex fluids disordered motions persist. 

While the first observation of such unconventional types of turbulence dates back to experiments in colloidal pipe flow in the early 20th century \cite{Ostwald-1926-KOLL}, elastic instabilities \cite{larson1990purely} and hence the potential driving mechanisms underlying such disordered motions were initially only discovered for a different class of flows, i.e., shear flows with streamline curvature. For such flow geometries, hoop stresses \cite{larson1990purely} arise and can be shown to render the laminar base flow linearly unstable. This mechanism is solely driven by elasticity, and instead of the Reynolds number, the relevant parameter is the Weissenberg number, $\mathit{Wi}=\lambda {\dot \gamma}$, the product of the polymer relaxation time,  $\lambda$, and the flow's shear rate, ${\dot \gamma}$. 
Accordingly, at a sufficiently high Weissenberg number, this instability can occur even in the zero inertia limit and support purely elastic turbulence, ET \cite{Groisman-2000-Nature, Steinberg-2021-ARFM}. However, being limited to curved flow geometries, this linear mechanism on its own cannot explain the broader phenomenology. Whether purely elastic turbulence can also arise in more common situations, such as in pipe flow, remains heavily debated \cite{Morozov-2005-PRL,Pan-2013-PRL,Samanta-2013-PNAS,Garg-2018-PRL,foggi2024unified,lellep_2024_purely,dubief2023elasto}. 

In their pipe flow experiments, by focusing on the nature of the transition to turbulence, Samanta et al.\cite{Samanta-2013-PNAS} provided a clear distinction between ordinary inertial turbulence and its viscoelastic counterpart. Since the latter disordered motions were observed at moderate but finite inertia, the authors dubbed this dynamical state elasto-inertial turbulence, EIT. The subsequent discovery of a linear instability for pipe flow \cite{Garg-2018-PRL, Chaudhary-2021-JFM} to a center mode traveling wave appeared to support the implied finite inertia requirement. For the Oldroyd-B polymer model used in their analysis, the instability ceases to exist at a lower inertia limit ($Re\approx63$). Subsequent experiments of pipe flow \cite{Choueiri-2021-PNAS}, however, found that the onset of EIT persisted to much lower $Re \approx 5$, supporting a possible subcritical route \cite{Wan-2021-JFM, Buza-2022-JFM}. At the same time, the characteristic structures observed in these experiments were shown to resemble the predicted center mode. For more elastic fluids, however, the instability was found to eventually disappear \cite{kamil_2026}, demonstrating that in this parameter regime the onset of EIT has a lower inertia limit, in qualitative but not quantitative agreement with the center mode stability result \cite{Garg-2018-PRL, Chaudhary-2021-JFM}. 

Unlike for pipes, the center mode instability in channel flow tends to persist to lower inertia levels. For ultra-dilute yet strongly elastic fluids \cite{Khalid-2021-PRL}, a regime that cannot be realized with available fluids in practice, the instability continues to the inertialess limit. For a more realistic polymer model with finite polymer extensibility (FENE-P model), exhibiting shear-thinning, the vanishing inertia limit could be confirmed for larger ranges of fluid parameters \cite{Buza-2022-JFM}. In direct numerical simulations of channel flow using the simplified Phan-Thien–Tanner (sPTT) constitutive model (again accounting for finite extensibility and shear-thinning), turbulent motions evolving around center mode structures could eventually be observed at vanishing inertia \cite{lellep_2024_purely} (Re of $10^{-2}$). 
Experiments of channel flow \cite{Pan-2013-PRL} reported chaotic motions at similarly low inertia levels, however, in this case the authors attributed the instability not to a linear instability, but to a subcritical variant of the aforementioned hoop stress instability, and initialization of turbulence required finite perturbation levels and curved streamlines. 

We will show in the following that viscoelastic turbulence persists in pipe flow experiments to the inertialess regime and that the motions have the chevron-type structures characteristic of the center mode instability. Consistent with the stability prediction \cite{Garg-2018-PRL}, the instability is initially found to diverge across a large range of fluid parameters. Eventually, however, for sufficiently high polymer concentrations and shear rates, the transition threshold drops sharply as the regime of purely elastic turbulence is entered. 

\begin{figure}[h]                                                                 
  \begin{center}
   \includegraphics[width=0.9\linewidth]{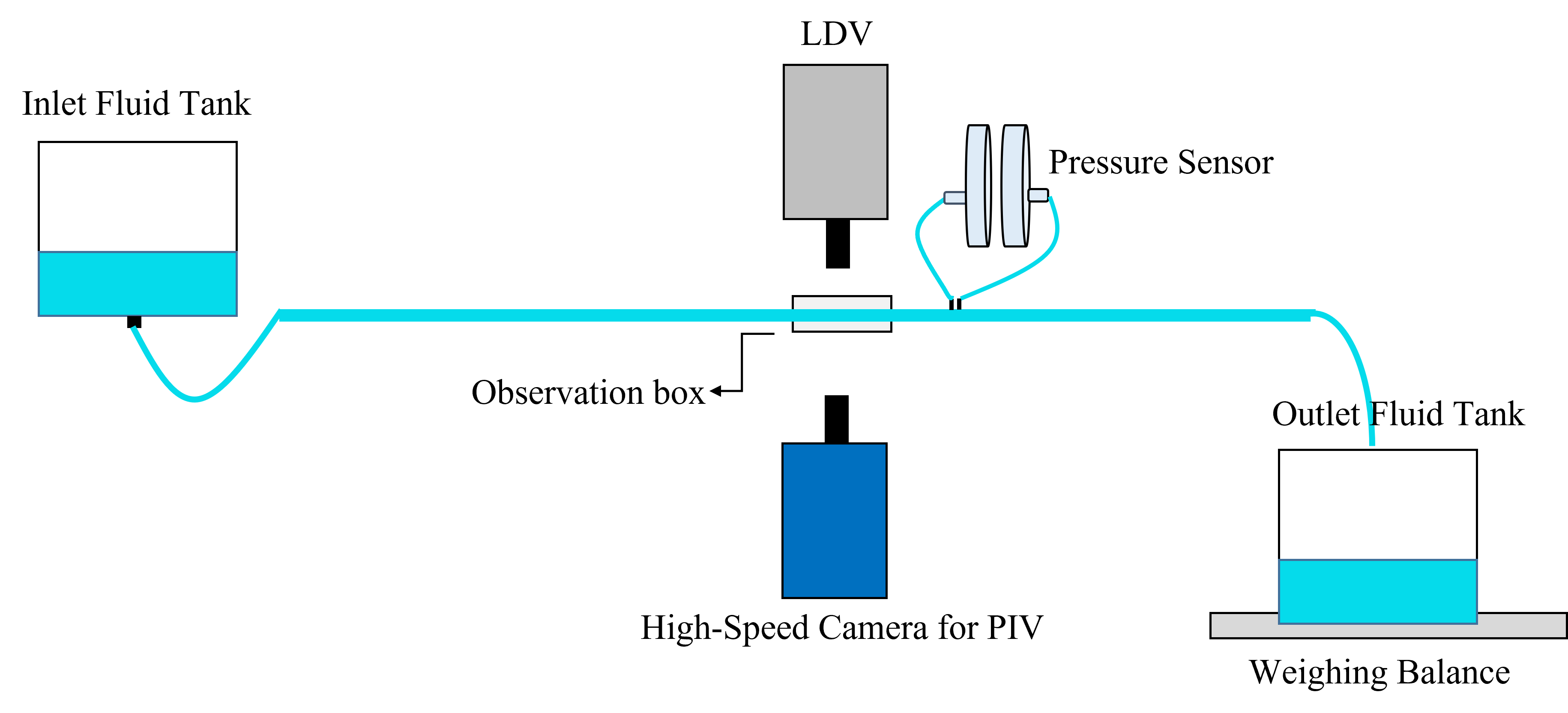}
  \end{center}  
  \caption{Schematic of the gravity-driven experimental facility with a pipe diameter of either 4 mm or 2 mm and, in each case, a length corresponding to $300D$. The fluid was supplied from an overhead reservoir, with the flow rate controlled by adjusting the hydrostatic head. Pressure measurements, particle image velocimetry (PIV), and laser Doppler velocimetry (LDV) were performed approximately $175D$ downstream of the pipe inlet to exclude entrance effects. The discharged fluid was collected over a prescribed time interval to determine the flow rate, which was subsequently used to calculate the Reynolds number.} 
\label{fig:setup}
\end{figure}

Experiments were conducted in two smooth borosilicate glass pipes with inner diameters of $2 \pm 0.01 mm $, and $4 \pm 0.01 mm $, each having a total length of $300 D$. A schematic of the setup is shown in Fig. \ref{fig:setup}. Flows were gravity-driven and, in the case of purely Newtonian liquids 
remained laminar up to $Re>2500$. 

As the working fluid, we used dilute solutions of a high molecular weight polymer (polyacrylamide of $18$ million Da) in water-glycerol mixtures. The viscosity ($\mu$) of the Newtonian solvent was varied by changing the glycerol content between $0-95\%$ by weight. It is noteworthy that, for a given polymer concentration, the relaxation time and hence the fluid's elasticity number, $E=Wi/Re$, is proportional to the solvent viscosity $\mu$ \cite{Choueiri-2021-PNAS}.

Dissolution of PAAm increases the viscosity, and the resulting polymer solutions are shear-thinning, i.e., the viscosity value decreases with shear rate. Both the shear-dependent viscosity and the longest polymer relaxation time were determined using an Anton Paar (MCR 102) rheometer. The Reynolds number was calculated using the shear-dependent viscosity evaluated for the actual wall shear rate ($\dot{\gamma}_w = 8U/D$), while the bulk velocity ($U$) was determined from the measured mass flow rate. The Weissenberg number (or elasticity number) and the ratio of the solvent to solution viscosity, $\beta$ (in our case computed based on the actual shear rate dependent viscosity), are then determined using the measured rheological properties. 

\begin{figure}[h]                                                                 
  \begin{center}
   \includegraphics[width=1\linewidth]{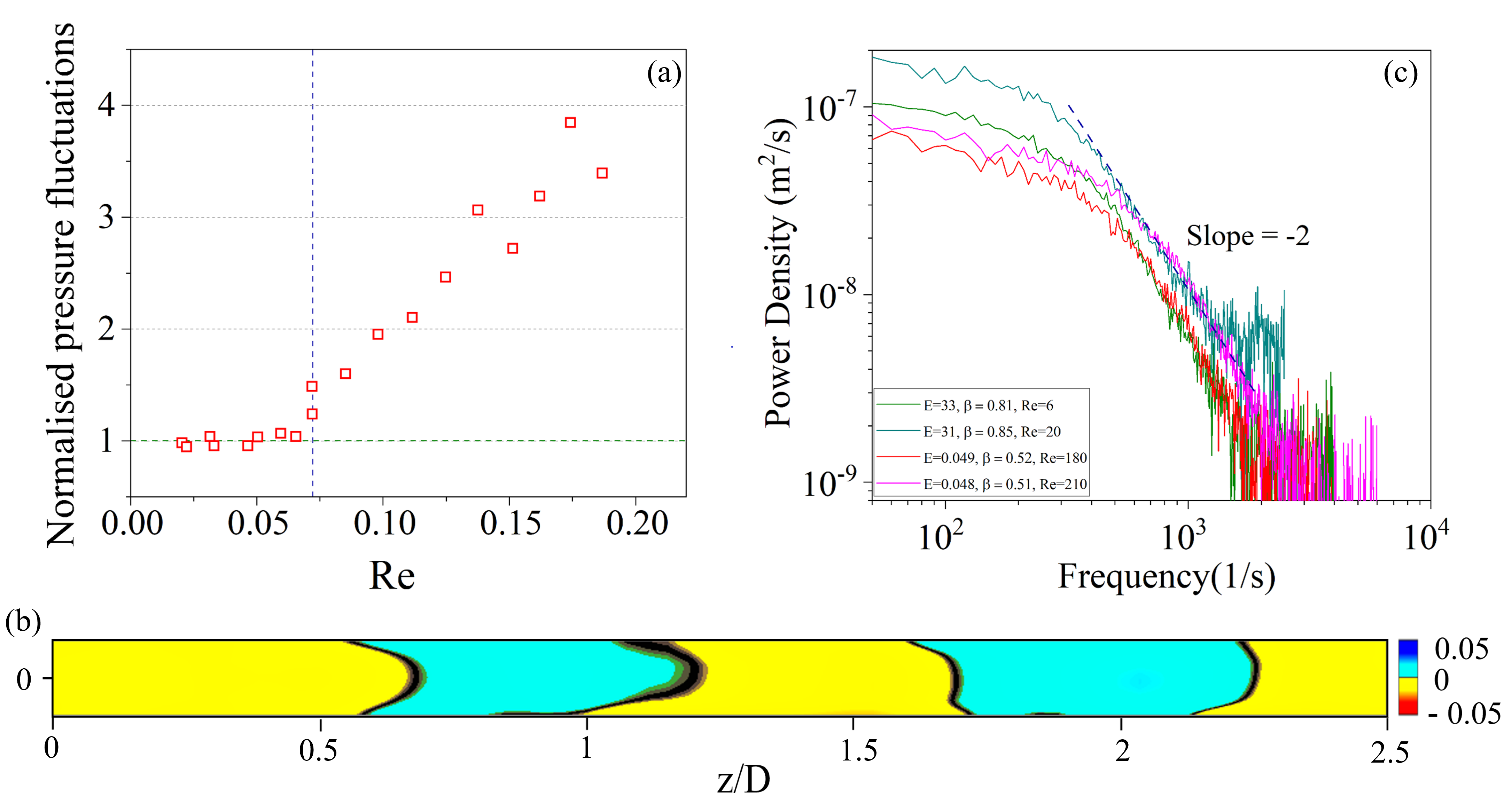}
  \end{center}  
  \caption{The Onset of elasto-inertial turbulence. (panel \textbf{a}): normalized pressure fluctuations marking the transition from the constant laminar (background noise) level to EIT. (panel \textbf{b}): Velocity field (measured using planar PIV) measured close to the onset of EIT. The colour scheme used is identical to that used by Choueiri et al. \cite{Choueiri-2021-PNAS} and highlights the relatively low flow amplitudes, marking high-speed regions in yellow and low-speed regions in green. Flows close to onset show the alternating chevron-type patterns characteristic of the center mode traveling wave at $E=20$ and $Re=3$. (panel \textbf{c}). Power spectra measured using LDV for various fluids above the onset of EIT. All data show the -2 scaling recently reported for the center mode close to onset \cite{lellep_2024_purely, lu2026multiple}.} 
\label{fig:Fig1}
\end{figure}

We conducted eight extensive experimental campaigns, six in the 4 mm pipe setup and two in the 2 mm setup, and in each case we selected a different polymer concentration (ranging from 20 to 100 ppm). For each fixed concentration, we mixed 15 to 20 solutions with different water to glycerol ratios (in the aforementioned range between 0 and 95\% glycerol content). For a given fluid, we then varied the Reynolds number and monitored the pressure drop across a short distance in the pipe. The onset of EIT is accompanied by a continuous rise in pressure fluctuations (see, e.g., \cite{Samanta-2013-PNAS, Choueiri-2018-PRL, Choueiri-2021-PNAS}). A representative example illustrating the threshold detection is shown in Fig. \ref{fig:Fig1}(a) (2 mm pipe, 100 ppm PAAm and 85\% glycerol), where each fixed Re measurement was conducted for an average of 5-15 minutes, and a fresh fluid sample was used in each case. 

Once the threshold was detected, we changed fluid parameters and repeated the full procedure, overall covering more than one hundred different polymer-water-glycerol combinations. Across the parameter space, the transition onset was found to be continuous and did not show any hysteresis, in agreement with earlier studies \cite{Samanta-2013-PNAS, Choueiri-2021-PNAS}. Moreover, close to onset flows are dominated by centre mode structures, an example is shown in  Fig. \ref{fig:Fig1}(b), and the velocity spectra show a $-2$ scaling, which has previously been reported for center mode driven turbulence \cite{lellep_2024_purely, lu2026multiple}.

\begin{figure}[h]                                                                 
  \begin{center}
   \includegraphics[width=1\linewidth]{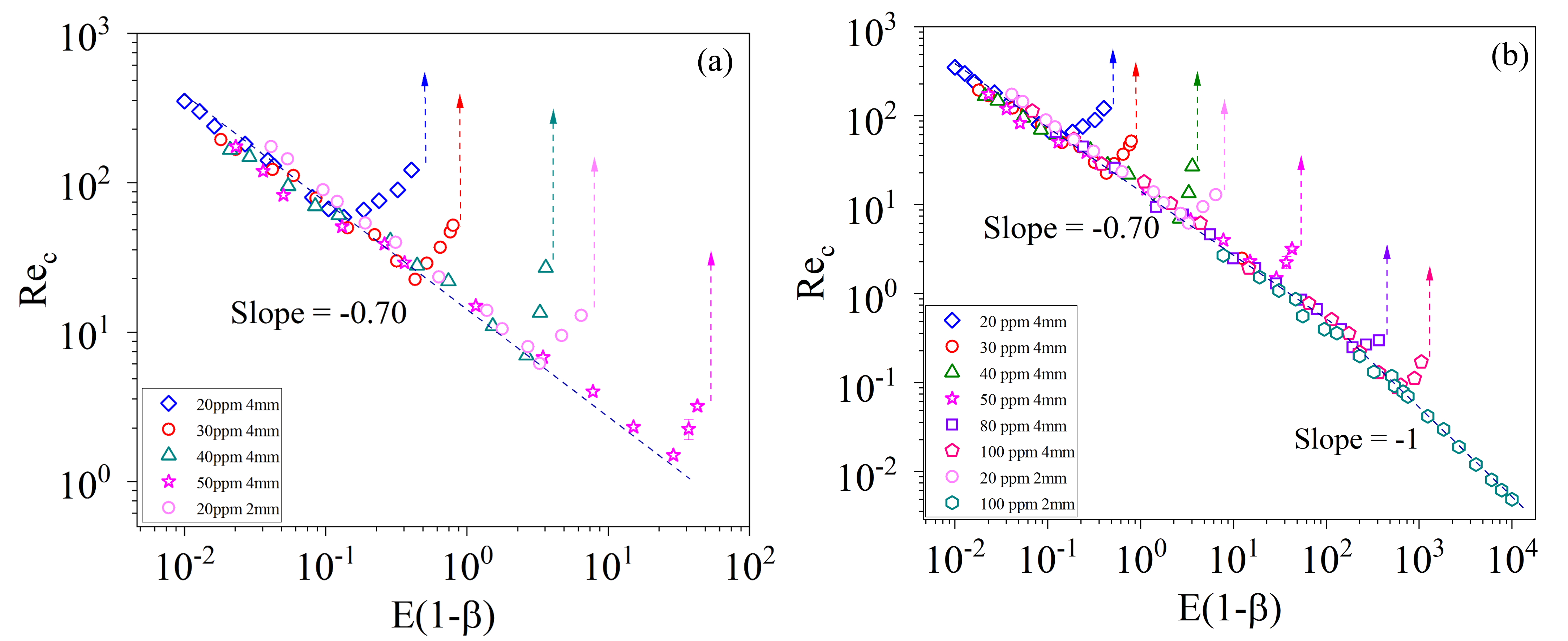}
  \end{center}  
  \caption{Scaling of the EIT transition threshold. (panel \textbf{a}): Onset of EIT in the elasto-inertial regime. As shown by the onset data, for the different polymer concentrations and for both pipes, collapses when plotted as a function of $E(1-\beta)$ and follows a power law of approximately $-0.7$ in agreement with \cite{kamil_2026}. For comparison, we include the 50 ppm data set from \cite{kamil_2026}. For all fluids tested in this regime, the onset of EIT eventually diverges. The divergence point moves to lower Re and higher elasticity number as the polymer concentration increases and as the pipe diameter decreases. Both increasing polymer concentration and decreasing pipe diameter increase shear-thinning.  (panel \textbf{b}): For polymer concentrations of 80 ppm and 100 ppm, the slope of the transition threshold steepens and approaches a value of -1, which marks the inertialess limit. While the data in the 4 mm pipe still diverge as this limit is approached, in the 2 mm pipe for a 100 ppm solution the transition thresholds now clearly follow the -1 scaling and no divergence can be found. The onset of EIT persists to vanishing inertia levels and can be traced in experiments to Re O($10^{-3}$).}
\label{fig:Fig2}
\end{figure}

Fluids were rheometrically analysed before and after measurements to ensure that no detectable degradation of the polymers had occurred. The results for the 20, 30, 40, and 50 ppm measurements are shown in Fig. \ref{fig:Fig2}(a). The data are presented in terms of the elasticity number $E=Wi/Re$, which is a fluid property, and hence, unlike $Wi$, $E$ does not change with flow rate. Moreover, as suggested in previous studies \cite{Chandra-2018-JFM, Chaudhary-2019-JFM, Khalid-2021-PRL, kamil_2026}, we selected the effective elasticity number $E(1-\beta)$, and we note that the different data sets collapse to a single curve with a slope of $\approx-0.7$, indicating an approximate power-law scaling as previously reported in \cite{kamil_2026}. While this slope differs from the predicted value of $-1.5$ for Oldroyd-B fluids, it is closer to the estimated value of $-5/8$  for FENE-P fluids \cite{Chaudhary-2021-JFM}. As shown by Kamil et al. \cite{kamil_2026}, for a given pipe diameter and polymer concentration, the instability threshold eventually diverges with increasing elasticity number, which is a clear signature of the center mode instability and hence in accordance with the linear stability result for pipe flow \cite{Garg-2018-PRL, Chaudhary-2019-JFM}. However, as mentioned above, the theoretical study predicted a lower bound for this center mode instability of $Re\approx63$, whereas in our experiments the onset of elasto-inertial turbulence reaches order of magnitude lower values, before the divergence occurs. As discussed in \cite{kamil_2026}, this quantitative discrepancy can be attributed to the oversimplified Oldroyd-B polymer model used for the stability analysis \cite{Garg-2018-PRL}. Two recent studies \cite{Morozov, Shankar} using more sophisticated polymer models (sPTT and FENE-P) that account for shear-thinning, observe that the center mode instability approaches the inertialess limit, in line with our experimental findings presented below. 

\begin{figure}[h]                                                                 
  \begin{center}
   \includegraphics[width=0.7\linewidth]{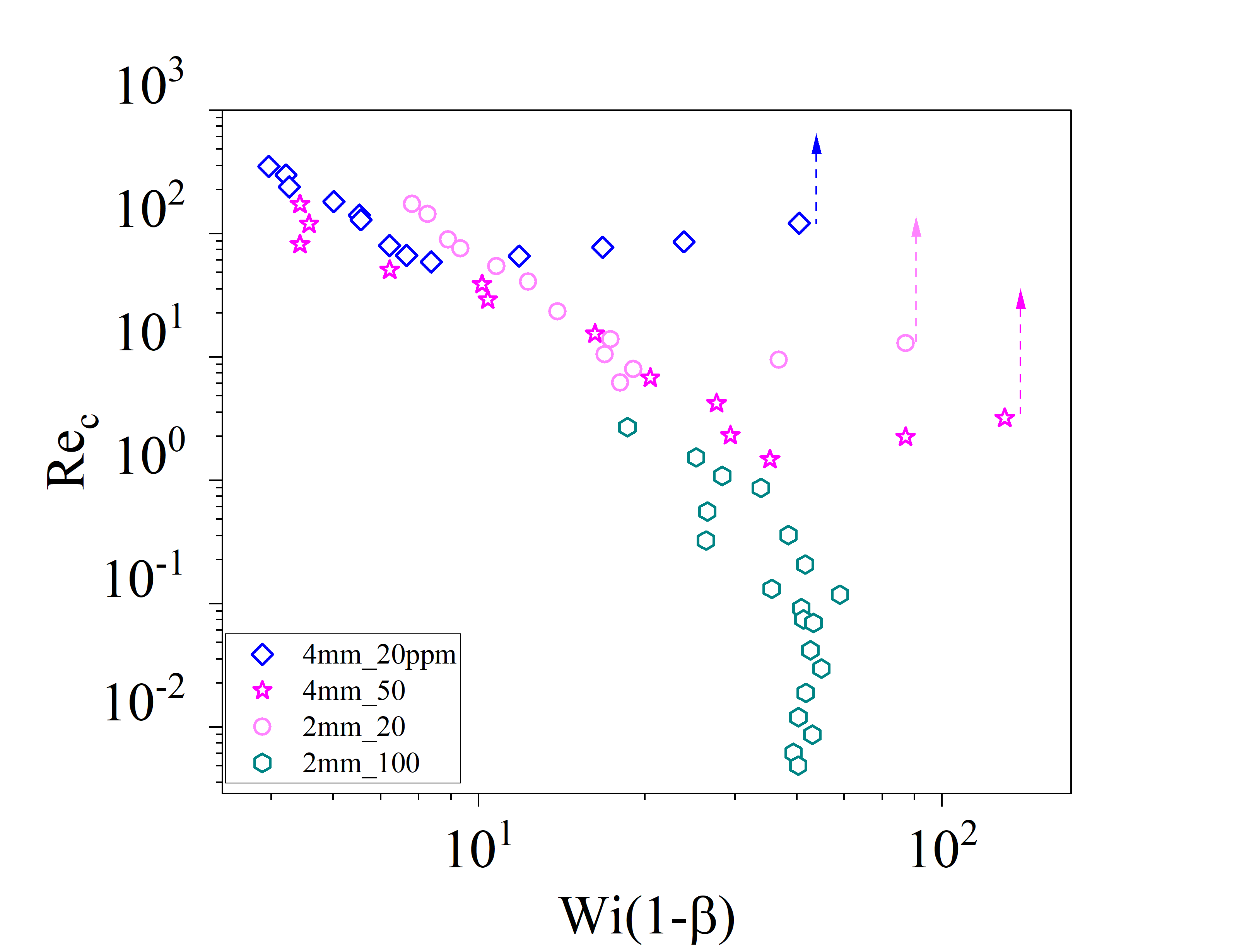}
  \end{center}  
  \caption{
 The Onset of the purely elastic regime. 
  Plotted as a function of $Wi(1-\beta)$, the transition threshold of the 100 ppm solution in the 2 mm pipe drops sharply at a constant Weissenberg number (green hexagons), marking the onset of the purely elastic regime. In contrast, for lower concentrations and in the larger diameter tube (blue, magenta and pink data sets), the instability threshold is limited to finite inertia levels and ceases to exist at an upper Weissenberg number limit (marked by the colored arrows). } 
\label{fig:Fig3}
\end{figure}

To extend the onset of EIT to lower inertia levels in our experiments, we increased the polymer concentration further to 80 and 100 ppm in the 4 mm pipe, which in turn increases the amount of shear-thinning exhibited by the respective fluid. As shown in Fig. \ref{fig:Fig2}(b), the transition threshold now begins to drop more quickly with $E(1-\beta)$ and is more closely approximated by a slope of -1. As we will discuss further below and as has been pointed out previously \cite{Khalid-2021-PRL}, in this parameter representation a -1 slope marks the inertialess limit. While in the crossover regime between these two scalings, we still find that the onset of EIT eventually diverges. This even applies to the 100 ppm solution, for which EIT persists to a minimum Re of order $10^{-1}$. To surpass this crossover regime and to further extend the -1 scaling, we switch to the 2 mm pipe where, due to the higher shear rates (four times higher than in the 4 mm pipe for a given Re), shear-thinning is much more pronounced, and we here chose a polymer concentration of 100 ppm. As shown by the green hexagons in Fig. \ref{fig:Fig2}(b), the data now fully fall onto the -1 slope, and no divergences could be found up to the lowest inertia levels that we could reach in our setup, corresponding to a $Re=4.9  \times 10^{-3}$. To illustrate that the -1 slope for $E(1-\beta)$ indeed corresponds to the inertialess regime, we replot the data in Fig. \ref{fig:Fig3} for the rescaled Weissenberg number ($Wi(1-\beta)$). While the data show much larger scatter for this representation, nevertheless, it becomes apparent that a -1 slope for the elasticity number corresponds to a constant threshold for the rescaled Weissenberg number, and hence the transition threshold ceases to diverge and instead sharply drops towards zero inertia. Our measurements suggest that beyond this constant $Wi(1-\beta)$ threshold, turbulence is purely elastic. Just like in ordinary fluids the transition to turbulence solely depends on the Reynolds number, here the transition and accordingly the state of fluid motion is governed by the Weissenberg number. 

The persistence of viscoelastic turbulence to zero inertia had first been anticipated a century ago \cite{Reiner-1926-1-KOLL} based on the conjecture that the instability would be solely driven by the shear rate and independent of the flow speed. Across a wide range of fluid parameters the transition scenario is however more complex and inertia remains relevant. Our findings suggest that the eventual approach to the inertialess regime does not only depend on the fluid's elasticity. The dependence on polymer concentration indicates that in addition shear-thinning, plays a key role.
\bmhead{Acknowledgements}
We thank V. Shankar and A. Morozov for valuable discussions. The technical support by the ISTA Machine Shop is gratefully acknowledged.
\bibliography{ref}
\FloatBarrier
\newpage

\end{document}